\documentclass{ws-ijbc}
\usepackage{ws-rotating}     % used only when sideways tables/figures are used
\usepackage{hyperref}
\begin{document}

\catchline{}{}{}{}{} % Publisher's Area please ignore

\markboth{Medeiros et al.}{An excitable electronic circuit as a sensory neuron model}

\title{An excitable electronic circuit as a sensory neuron model}

% \title{``This paper is submitted to the Special Issue of 2010DDays''
% \\ \ \\
% An excitable electronic circuit as a sensory neuron model}

\author{Bruno N. S. Medeiros$^{1}$, Victor Minces$^2$, Gabriel
  B. Mindlin$^3$, Mauro Copelli$^1$, Jos\'e R. Rios Leite$^1$}
\author{}
\address{\vspace{15pt}$^1$Departamento de F\'isica, Universidade
  Federal de Pernambuco, 50670-901, Recife, PE, Brazil.}
\author{}
\address{$^2$Department of Cognitive Neuroscience, University of California San Diego, 92093-0515, La Jolla, CA, USA}
\address{$^3$Departamento de F{\'\i}sica, FCEN, Universidad de Buenos
  Aires, Ciudad Universitaria, Pab. I, 1428, Buenos Aires, Argentina}

% \author{Bruno N. S. Medeiros$^1$}
% \address{$^1$Departamento de F\'isica, Universidade Federal de Pernambuco, 
% Recife, Pernambuco, Brazil\\
% %Av. Prof. Moraes Rego, 1235, Cidade Universit\'aria\\
% %Recife, Pernambuco CEP: 50670-901, Brazil\\
% br.medeiros@gmail.com}

% \author{Victor Minces$^2$}
% \address{Group, Company, Address\\
% City, State ZIP/Zone, Country\\
% sauthor@company.com}

% \author{Gabriel B. Mindlin}
% \address{}
% \author{Mauro Copelli}
% \address{}
% \author{Jos\'e R. Rios Leite$^1$}
% \address{}

\maketitle

\begin{history}
\received{(to be inserted by publisher)}
\end{history}

\begin{abstract}
  An electronic circuit device, inspired on the FitzHugh-Nagumo model
  of neuronal excitability, was constructed and shown to operate with
  characteristics compatible with those of biological sensory
  neurons. The nonlinear dynamical model of the electronics
  quantitatively reproduces the experimental observations on the
  circuit, including the Hopf bifurcation at the onset of tonic
  spiking. Moreover, we have implemented an analog noise generator as
  a source to study the variability of the spike trains. When the
  circuit is in the excitable regime, coherence resonance is observed.
  At sufficiently low noise intensity the spike trains have Poisson
  statistics, as in many biological neurons. The transfer function of
  the stochastic spike trains has a dynamic range of $6$~dB, close
  to experimental values for real olfactory receptor neurons.
\end{abstract}

\keywords{Electronic circuit, Excitable element, Coherence resonance,
  Dynamic range.}

%\begin{multicols}{2}
\section{Introduction}

Ever since the pioneering work of~\citet{HH52d}, the biophysical
mechanisms underlying the generation and propagation of action
potentials (spikes) in neurons have been described with increasing
detail, ranging from the discovery of new types of ion channels to the
study of intracellular calcium dynamics~\cite{Hille}. No matter how
interesting, these new findings have helped little in our
understanding of collective neuronal phenomena, which remain a
daunting task in face of the interplay among high-dimensionality,
noise and nonlinearity (see e.g.~\citet{Chialvo10a} for a recent
review). The challenge should nonetheless be faced: the solution of
issues at the frontiers of current-day neuroscience, like
e.g. grandmother cell~\cite{Barlow72} versus population
coding~\cite{Young92}, or firing rate versus spike-time
coding~\cite{Spikes} will likely be grounded on our success in this
endeavor.

In fact, theoretical progress in this front has been achieved in
recent years with very simple models. One such example is the proposed
solution for the century-old problem of the origin of psychophysical
response curves~\cite{Copelli02,Copellinatphy}. Steven's
psychophysical law states that the psychological perception $F$ of a
physical stimulus (e.g. light, or odorant) of intensity $h$ is a power
law $F \propto h^s$, with experimental values of the Stevens exponent
$s$ fluctuating around $s\simeq 0.5$. Compared to a linear response,
psychophysical nonlinear responses have at least one evolutionarily
favorable property: they amplify weaker stimulus, i. e. they have a
larger dynamic range. But how do the Stevens exponents arise in the
nervous system?

At first, this question seems puzzling because single neurons
typically have small dynamic ranges~\cite{Rospars00}. A theoretical
solution recently proposed involves a collective phenomenon: excitable
waves are generated by the incoming stimuli and propagate
``laterally'' among excitable neurons, thereby amplifying the system
response (in comparison to what would be observed in the absence of
the coupling). Interestingly, this amplification mechanism is
self-limited: under intense stimulation, for instance, a large number
of excitable waves can be created, but owing to refractoriness they
annihilate upon collision. The enhancement of dynamic range in this
model is therefore governed by the low-stimulus
amplification~\cite{Copelli02,Copellinatphy}. Robustness of these
results has been tested at different modeling
levels~\cite{Copelli05a,Ribeiro08a,Assis08,Publio09}, showing that the
degree of biophysical realism in the model of each neuron is less
relevant to the global dynamics than the topology of the
network~\cite{Copelli05b,Copelli07,Ribeiro08a,Assis08,Gollo09}. This
phenomenon has also been studied
analytically~\cite{Furtado06,Larremore11} and was recently confirmed
experimentally in cortical slices~\cite{Shew09}.

The appeal of a sensory system with large dynamic range based on a
network of simple excitable units, each with small dynamic range, goes
beyond basic research in neuroscience. The idea could be reversed,
leading to biologically inspired artificial sensors, which have been
used in a variety of scenarios~(see e.g.~\citet{Celso99}). 

\begin{figure}[tb]
  \centering
  \includegraphics[width=0.5\linewidth]{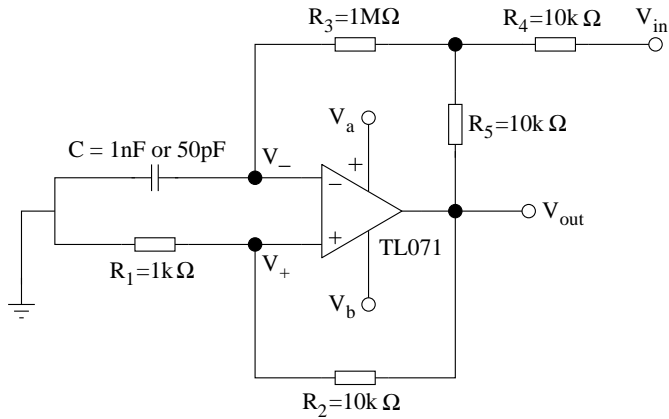}
  \caption{Excitable electronic circuit. $V_a$ and $V_b=-V_a$ are the
    operational amplifier supply voltages. $V_{in}$ is an input
    voltage, corresponding to an external stimulus. We describe the
    circuit as a two-dimensional dynamical system on the variables
    $V_-$ and $V_{out}$ (see Eqs.~\eqref{eq:sis}).}
  \label{fig:circuit}
\end{figure}

There are several electronic circuits reported in the literature which
have been designed to present a neuron-like dynamical response. The
rationale behind those efforts was to dynamically interact with
biological neurons rather than stimulating them using response
independent current commands. In this way, electronic circuits which
analogically integrated the Hindmarsh and Rose equations~\cite{Szucs00}
were coupled to the neurons of a preparation of lobster pyloric CPG
neurons. This allowed to show that regularity could emerge as a
collective dynamical property of units which individually presented
complex dynamics. In another set of experiments, electronic neurons
interacting with a biological preparation were used to unveil which
dynamical properties of a neural network depend on the bifurcation
leading to excitation for the units, rather than on the details of the
neural dynamics. To carry out this program, a standard form for class
I excitable dynamics was analogically integrated with a circuit, which
was used to replace a neuron in a midbody ganglion of the leech Hirudo
medicinalis~\cite{Aliaga03}. The responses under the stimulation of both the
natural preparation and the one with a replaced neuron were found to
be similar. Beyond the possibility of interacting with neurons through
a dynamically sensible way, these efforts provide empirical support to
the program of studying neural processes through simple and relatively
low dimensional dynamical systems.  Depending on the question under
study, it might be desirable to be able to establish a closer link
between the device and a neuron. In this spirit, a device implementing
a conductance model was recently proposed~\cite{Sitt07}.

These circuits, however, have two limitations for our purposes. First,
they are still too complex to be replicated in large scale. Second,
they do not have a controllable noise source to produce stochastic
spike trains, a feature that is common to the both
models~\cite{Copelli02} and real
neurons~\cite{Abott,MainenSejnowski95,Petracchi95}. The present work is
a first step in this direction. We propose an excitable electronic
circuit which can serve as a building block of an electronic
sensor. The advantages of its extreme simplicity are twofold: it
allows for scalability and, at the same time, simple mathematical
modeling.

The paper is organized as follows. In section~\ref{model}, we describe
the electronic circuit and the equations that model its dynamics. In
section~\ref{noise}, we introduce noise from a simple analog noise
generator at the input of the excitable circuit and study the
statistical properties of the resulting spike trains and show that it
can exhibit Poisson statistics as well as coherence resonance, as
expected. In section~\ref{range} we evaluate the dynamic range of the
excitable circuit and show that it is comparable to that of single
sensor neurons.

\section{\label{model}Dynamic model}
The circuit we propose is shown in Fig.~\ref{fig:circuit}. It is
composed of five resistors, one capacitor and one operational
amplifier. The voltage $V_{in}$ corresponds to an external stimulus,
which can be e.g. a constant or the sum of DC and noise voltages. In
our electronic neuron, the operational amplifier behaves as a simple
comparator circuit, for which we use the following nonlinear model:
\begin{equation}
  \label{eq:compmodel}
  \frac{dV_{out}}{dt}=S\,\mbox{sign}\left[V_b-V_{out}+(V_a-V_b)\Theta(V_+-V_-)\right], 
\end{equation}
where $\Theta$ is the Heaviside function and $S$ is the op-amp slew
rate (whose datasheet value for the simple TL071 in the circuit is $S=
16$~V/s). As usual, symmetric supply voltages $V_b=-V_a$ were used.

\begin{figure}[t]
    \centering  
  \includegraphics[width=.7\linewidth]{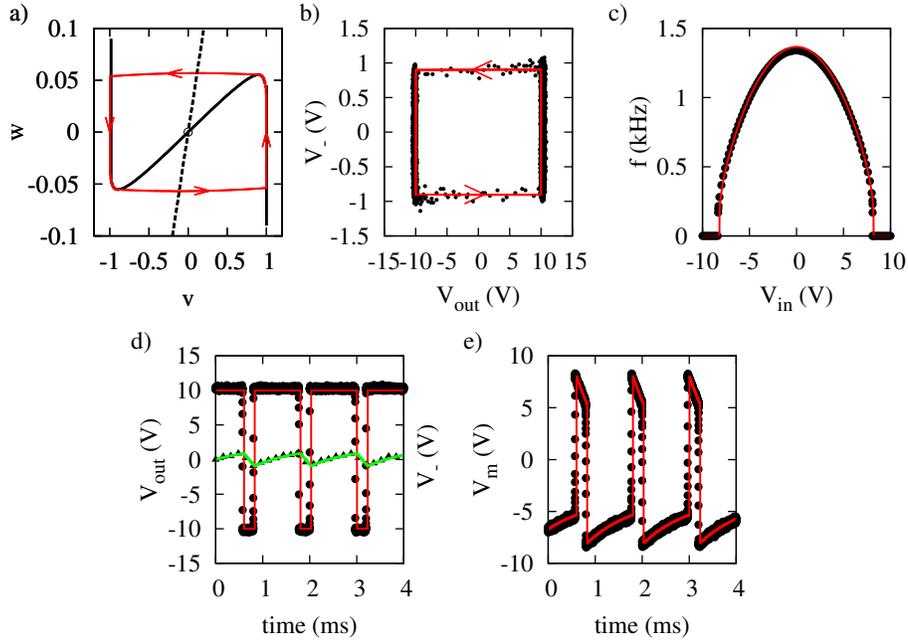}
  \caption{a) Nullclines of system \eqref{eq:adsis} for $a=1$, $b=1$,
    $\alpha=0.0909$, $\beta=0.5$, $\gamma=0.5$, $j=0$, $\phi=0.01$ and
    $x_0=9\times 10^{-3}$: solid black line for the $\dot v=0$
    nullcline and dashed black line for the $\dot w=0$ nullcline. The
    fixed point is unstable and the trajectories are attracted to a
    limit cycle (red solid line). b) Experimental limit cycle (black
    dots) and numerical integration of the model (red solid line) for
    $x_0=1\times 10^{-5}$, $V_a=10$~V, $V_b=-10$~V, $V_{in}=-6$~V and
    $\phi =5 \times 10^{-4}$ (other parameters are the same as in
    (a)). c) Experimental frequency response $f$ to the external DC
    stimulus $V_{in}$ (black dots) and the same for the numerical
    integration of the model (red line).  d) Comparison between
    experimental time series of $V_{out}$ and $V_-$ (black circles and
    triangles, respectively) with numerical integration of the model
    (red and green lines, respectively). e) Experimental (black dots)
    and numerical (red line) spike trains obtained from the analog
    subtraction $V_m$ of the dynamical variables (see text for
    details).}
  \label{fig:nonlin-exp}
\end{figure}

Assuming $R_3\gg R_4,R_5$ and applying Kirchhoff's laws, we arrive at
a two-dimensional dynamic model on the variables $V_{out}$ and $V_-$:

\begin{subequations}
  \label{eq:sis}
\begin{eqnarray}
\frac{dV_{out}}{dt} &=& \frac{V_c}{\varepsilon}\,\mbox{sign}\left[V_b-V_{out}+
    (V_a-V_b)\Theta(\alpha V_{out}-V_-)\right]\; , \\
\frac{dV_-}{dt}&=&\frac{1}{R_3C}\bigl[\beta V_{out}+\gamma V_{in}-V_-\bigr]\; .
\end{eqnarray}
\end{subequations}
where $\alpha\equiv R_1/(R_1+R_2)$, $\beta\equiv R_4/(R_4+R_5)$ and
$\gamma\equiv R_5/(R_4+R_5)$. $V_c=10$~V is a characteristic voltage
of the same order of magnitude of the supply voltages, and we have
defined $\varepsilon\equiv V_c/S$ as a characteristic (short) time
scale. To avoid the possibility that the system~\eqref{eq:sis} has
more than one fixed point, we require $\beta>\alpha$.  In terms of the
variables

\begin{subequations}
  \label{eq:vw}
  \begin{eqnarray}
   v  & \equiv &\frac{V_{out}}{V_c}\; ,  \\
   w & \equiv & \frac{V_{-}}{V_c}\; ,
  \end{eqnarray}
\end{subequations}
the equations can be rewritten in dimensionless form
\begin{subequations}
\label{eq:adsis}
\begin{eqnarray}
\dot{v} &=& \mbox{sign}\left(b-v+\frac{(a-b)}{1+e^{-(\alpha v-w)/x_0}}\right)\; , \\
\dot{w} &=& \phi \left[\beta v+\gamma j-w\right],
\end{eqnarray}
\end{subequations}
where we defined the dimensionless groups:
\begin{equation}
  \tau \equiv \frac{t}{\varepsilon}\; 
  ;~a=\frac{V_a}{V_c}\; 
  ;~b=\frac{V_b}{V_c}\;
  ;~\phi=\frac{\varepsilon}{R_3C}\;
  ;~j=\frac{V_{in}}{V_c},
\label{gr-ad}  
\end{equation}
and replaced $\Theta$ by the continuous function 
\begin{equation}
  \label{eq:thetatilde}
  \tilde{\Theta}(x;x_0) = \frac{1}{1+e^{-x/x_0}}
\end{equation}
for the purpose of numerical integration and derivation (see
below). Note that $\tilde{\Theta}\to\Theta$ as $x_0\to 0$. The
constant $\phi\ll 1$ sets the ratio between the fast and slow time
scale as in the FitzHugh-Nagumo model, so that $R_3C$ ultimately
controls the overall time scale of the problem.

As shown in Fig.~\ref{fig:nonlin-exp}a (black lines), the nullclines
$\dot v=0$ and $\dot w =0$ of Eqs.~\eqref{eq:adsis} resemble those of
the FitzHugh-Nagumo model for neuronal excitability, with one fast
($v$ or $V_{out}$) and one slow ($w$ or $V_-$) variable. In the limit
$x_0\to 0$, the cubic-like $\dot v=0$ nullcline becomes piecewise
linear. When the fixed point sits at its outer branches, it is
stable. It loses stability in a Hopf bifurcation as the $w$ nullcline
crosses the $v$ nullcline at its central branch, so trajectories are
attracted to a limit cycle (red line) with nonzero frequency $f$
(i.e. $f$ changes discontinuously at the bifurcation). Below the Hopf
bifurcation, the circuit is said to be type-II
excitable~\cite{Rinzel98b}. 

\begin{figure}[t]
  \centering  
  \includegraphics[width=.7\linewidth]{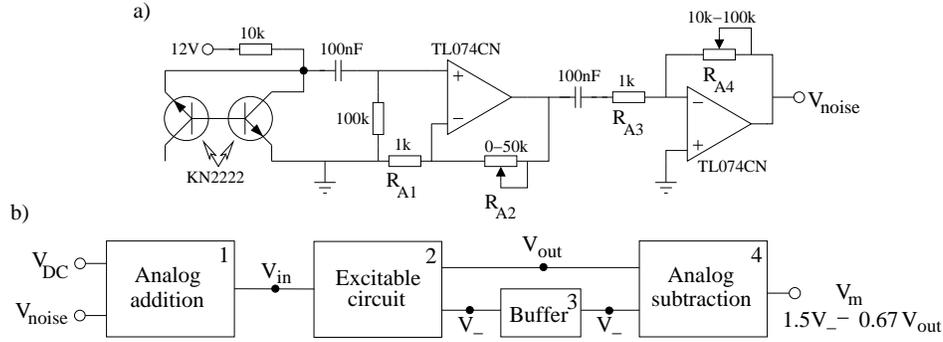}
  \caption{a) An analog noise generator based on the amplification of
    transistors thermal noise. Noise amplification is given by
    $A=[(R_{A1}+R_{A2})/R_{A1}](R_{A4}/R_{A3})$.  b) Block diagram of
    the circuit used to verify the excitability of the circuit
    presented in Fig.~\ref{fig:circuit}. Analog addition and
    subtraction (see text for details) are performed with standard
    TL074 op-amp operations~\cite{Senturia}.}
  \label{fig:block}
\end{figure}

There is good quantitative agreement between experimental data from
the circuit and the numerical integration, as can be seen in
Fig.~\ref{fig:nonlin-exp}b, c, d and e. Note that through an analog
subtraction $V_m \equiv 1.5V_--0.67V_{out}$ (see also
Fig.~\ref{fig:block}b) the circuit exhibits the spikes typical of
neuronal membrane potentials (Fig.~\ref{fig:nonlin-exp}e). We
emphasize that in Fig.~\ref{fig:nonlin-exp} experimental and
numerical data agree without any fitting parameter, as long as $x_0$
is sufficiently small ($\lesssim 10^{-4}$).

\section{\label{noise}Noise addition and coherence resonance}

So far we have discussed the response of the excitable circuit under
DC stimulation. Biological neurons, however, can show highly variable
responses, even when subjected to a presumably constant
stimulus. Examples range from highly variable responses olfactory
receptor neurons (ORNs) to presentation of identical puffs of
odorants~\cite{Rospars00}, to cortical cells stimulated with a
constant current via an intracellular
electrode~\cite{MainenSejnowski95}. In an attempt to endow our excitable
circuits with the variability in the spike trains observed in
biological neurons, we propose the simple analog noise generator shown
in Fig.~\ref{fig:block}a. Once more, its simplicity allows one to
attach independent noise generators to each excitable circuit when
connecting them in a network.

\begin{figure}[t]
  \centering  
  \includegraphics[width=.7\linewidth]{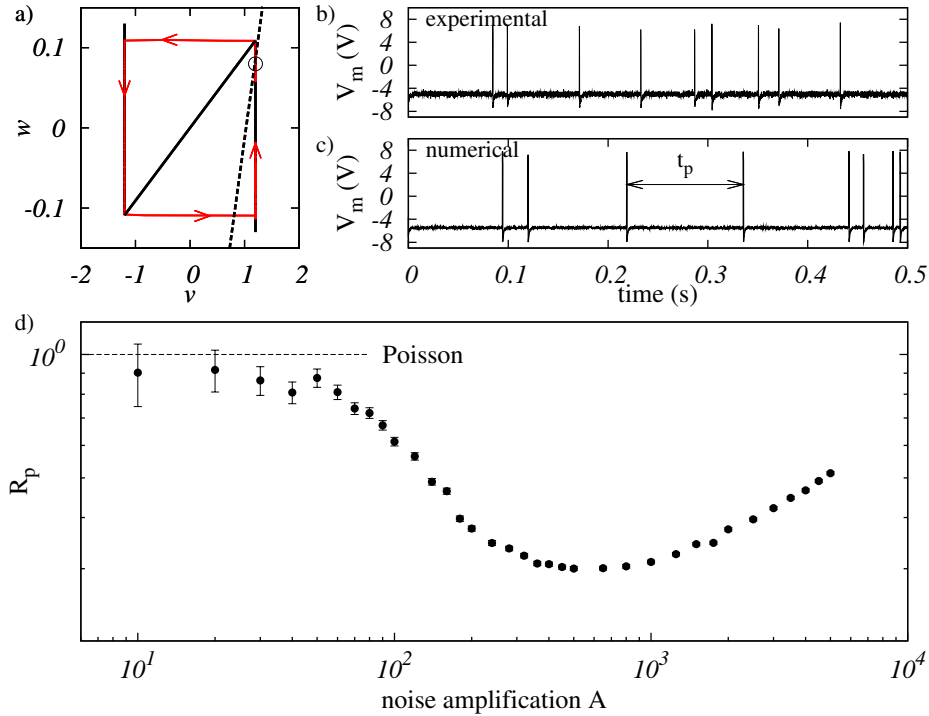}
  \caption{a) Numerical phase plane trajectory (red line) due to noise
    excitation. Without noise the system would stay in a resting state
    at the fixed point (white dot). Experimental (b) and numerical (c)
    spike train series are shown when the system is in the excitable
    state (stable fixed point as shown in (a)). d) Experimental
    coherence resonance curve for $C=50$~pF, $V_a=12~\mbox{V}=-V_b$
    and $V_{DC}=-7.826$~V (see Fig.~\ref{fig:circuit}). Each point
    corresponds to an average over 10~s time series.}
  \label{fig:cohe}
\end{figure}

The circuit in Fig.~\ref{fig:block}a provides a two-stage
amplification control via two operational amplifiers to the thermal
noise produced by the KN2222 transistors. Its output voltage
$V_{noise}$ is approximately a Gaussian white noise voltage with a
cutoff frequency around $1$~kHz.

To obtain variable spike trains, the stimulus $V_{in}$ consists in the
analog addition of $V_{DC}$ and $V_{noise}$ (see blocks 1 and 2 in
Fig.~\ref{fig:block}b). In the model, this corresponds to replacing
Eq.~(\ref{eq:adsis}b) with
\begin{equation}
  \label{eq:sisnoise}
\dot{w} = \phi \left[\beta v+\gamma j  + D\xi(t) -w\right],
 \end{equation}
 where $D$ grows linearly with the gain in the noise amplification $A$
 (which in turn is controlled by the variable resistors shown in
 Fig.~\ref{fig:block}).

Setting $V_{DC}$ below the Hopf bifurcation, the circuit sits at a
stable fixed point at the right branch of the $\dot v=0$ nullcline,
from which it eventually departs owing to noise
(Fig.~\ref{fig:cohe}a). This generates spike trains with variable
interspike intervals $t_p$, as shown in Fig.~\ref{fig:cohe}b and c.

We now show that the interplay between noise and excitability behaves
as expected in our simple circuits. \citet{Kurths97} 
have shown that the coherence the spike train of the
FitzHugh-Nagumo model peaks at an intermediate noise value, in a
phenomenon which has been called ``coherence resonance''. In other
words, the normalized standard deviation
\begin{equation}
  \label{eq:rp}
  R_p\equiv \frac{\sqrt{\langle
    t_p^2\rangle-\langle t_p\rangle ^2}}{\langle t_p \rangle}
\end{equation}
should have a minimum as a function of the noise intensity. This is
precisely what we observe in our circuit when $V_{DC}$ ($=-7.826$~V)
is close to the Hopf bifurcation ($V_{Hopf}=-7.82$~V), as displayed in
Fig.~\ref{fig:cohe}d. Note that $R_p$ close to zero means that the
time series is approximately periodic.

\begin{figure}[b]
  \centering  
  \includegraphics[width=.7\linewidth]{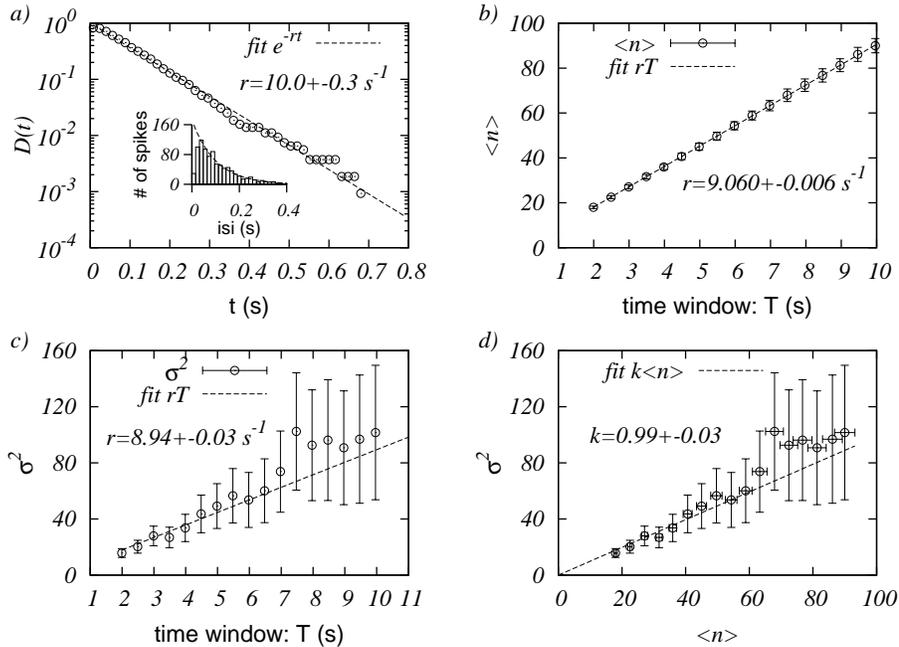}
  \caption{Spike train statistics of a 100 s duration series from
    the circuit of Fig.~\ref{fig:block}b. a) Interspike interval (isi)
    accumulated distribution in log-linear scale. Inset: corresponding
    histogram of isi. The dashed line corresponds to an exponential
    fit of a Poisson distribution with mean firing rate $r=10.0(3)$
    s$^{-1}$. The deviation from the Poisson distribution for small isi
    is due to the refractoriness of the excitable
    circuit. In the following graphs we have divided the series in
    time windows of duration $T$. The mean number of spikes
    $\left<n\right>$ (b) and the variance $\sigma ^2$ (c) are shown as
    functions of $T$. In (d) we have $\sigma ^2$ as a function of
    $\left<n\right>$. The dashed lines are fits of $\sigma
    ^2=\left<n\right>=rT$ according to the Poisson distribution.}
  \label{fig:isi}
\end{figure}

For small noise amplitudes ($V_{noise}\sim 50$~mV, or $A\sim {\cal
  O}(1)$ in Fig.~\ref{fig:cohe}d), spikes are sparse and $R_p$
approaches unity. This suggests a Poisson process in which the
interspike interval distribution approaches an exponential
\begin{equation}
  \label{eq:poi}
  P(t_p)=re^{-rt_p},
\end{equation}
where $r$ is time rate constant. This Poisson limit is interesting
because it is observed in different neuronal
preparations~\cite{Abott,Petracchi95}, so we performed a
detailed statistical analysis of the small $V_{noise}$ regime.

In Fig.~\ref{fig:isi}a the statistics of a $100$~s experimental time
series was compared to the accumulated distribution
\begin{equation}
  \label{eq:dpoi}
  D(t)\equiv \int^{\infty}_tre^{-rt_p}dt_p=e^{-rt} ,
\end{equation}
showing good agreement for a fitted rate $r\simeq
10.0(3)$~s$^{-1}$. To check for consistency, we divided the time
series in small time windows of size $T$ and sampled the number $n$ of
of spikes per window. In a Poisson process one has the linear
relationships $\langle n \rangle = rT$, $\sigma_n^2 \equiv \langle n^2
\rangle - \langle n \rangle^2 = rT$ which are confirmed
in~Fig.~\ref{fig:isi}b and c. The unit slope in the $\sigma_n^2$
versus $\langle n \rangle$ plot is also verified (see
Fig.~\ref{fig:isi}d). These results show that our circuit can be used
to mimic not only deterministic dynamics, but also simple statistical
properties which appear in biological neurons.

\section{\label{range}Dynamic range}

In this section we study the response of our excitable system to
varying input voltage $V_{DC}$, considering the noise amplitude
$V_{noise}$ constant. Although in real neurons the background noise
may have a dependence on the stimulus, it is a fair approximation to
treat the noise amplitude as constant and focus on the dependence on
input signal as a control parameter of the dynamics. In what follows,
the response of the circuit is defined as the mean firing rate $F$
measured over a fixed time interval $T_{m}$. This so-called ``rate
coding'' is also a longstanding approximation~\cite{Adrian26}, which
seems  to fit data in several cases~\cite{Koch,Arbib02}.

\begin{figure}[!b]
  \centering  
  \includegraphics[width=.7\linewidth]{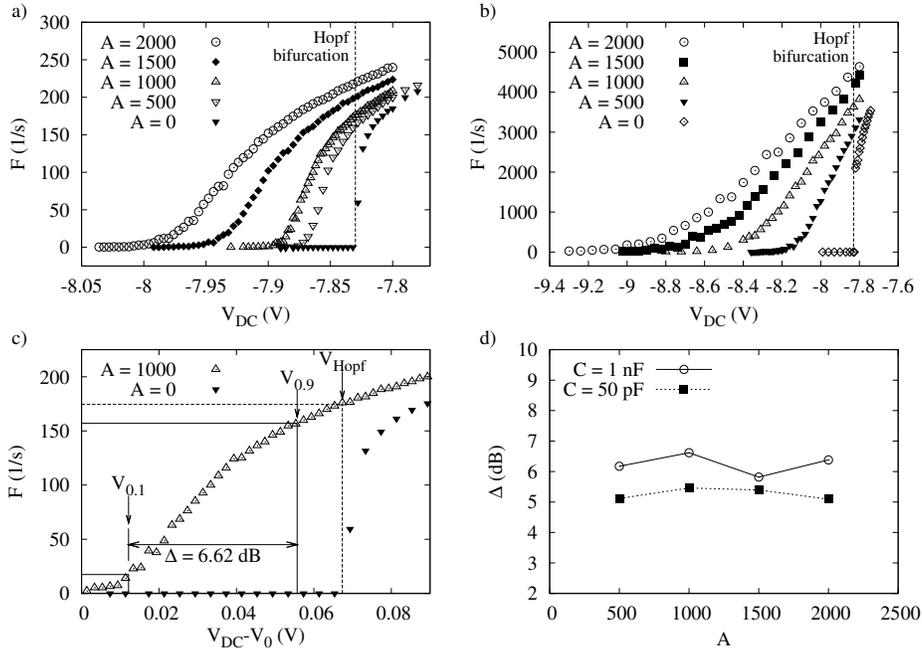}
  \caption{Experimental response curves $F(V_{DC})$ measured at
    different values of the noise amplification $A$. Supply voltages
    $V_a = 12~\mbox{V}= -V_b$. a) $C=1$~nF ($\phi =5\times 10^{-4}$
    and $T_m=10$~s).  b) $C=50$~pF ($\phi =0.01$ and $T_m=0.2$~s). c)
    Response curve for $C=1$~nF (A=1000), and relevant parameter for
    calculating the dynamic range. d) Dynamic range as function of
    noise amplification for $C=50$~pF (black squares) and $C=1$~nF
    (white circles).}
  \label{fig:dynamic}
\end{figure}

For fixed $T_m$ and noise amplification $A$, the response $F$ of our
circuit is an increasing function of the stimulus $V_{DC}$ because
larger values of $V_{DC}$ amounts to increased excitability, lowering
the ``effective threshold'' to noise-induced spike generation (there
is no real threshold in type-II excitable
neurons~\cite{Izhikevich}). Conversely, for fixed $V_{DC}$, the
response $F$ also increases with increasing noise intensity $A$. These
results are shown in Fig.~\ref{fig:dynamic}a, where we plot (for
different noise intensities) the responses $F(V_{DC})$ of our
excitable circuit with a 1~nF capacitor. This choice sets the time
scale of the neuron in the millisecond range (i.e. that of biological
neurons). Note that in the absence of noise ($A=0$) the response is
null up to the Hopf bifurcation (so the lowest curve in
Fig.~\ref{fig:dynamic}a is similar to Fig.~\ref{fig:nonlin-exp}c). 

Results in Fig.~\ref{fig:dynamic}b correspond to a circuit with a
50~pF capacitor. This single change renders a much faster circuit, now
operating in the microsecond range, but with its dynamical features
otherwise preserved. This has potential applications, because a faster
circuit requires shorter measurement intervals $T_m$ (=0.2~s in our
example) for a reliable estimation of the firing rate. 

Given a response curve, we can calculate its dynamic range, which
roughly speaking corresponds to the range of stimulus intensity that
the firing rate can ``appropriately code''. Measured in decibels, this
is arbitrarily defined as~\cite{Rospars00,Copelli05b}
% \begin{equation}
%   \label{eq:dynamic}
%   \Delta \equiv 10\log_{10}\left(\frac{V_{0.9}^\prime}{V_{0.1}^\prime}\right),
% \end{equation}
\begin{equation}
  \label{eq:dynamic}
  \Delta \equiv 10\log_{10}\left(\frac{V_{0.9}^*}{V_{0.1}^*}\right),
\end{equation}
where $V_x^* \equiv V_x - V_0$ are measured relative to the
voltage $V_0$ at which the response becomes non-zero and
\begin{equation}
F(V_x)=xF_{max} \ \ \ \ \ \ (0\leq x \leq 1) \; ,
\end{equation}
where $F_{max}$ is the firing rate at the Hopf bifurcation. In words
(see Fig.~\ref{fig:dynamic}c), $\Delta$ measures the range of stimulus
$V_{DC}$ which are neither too small ($V_{DC} < V_{0.1}$) to go
undetected nor too close ($V_{DC} > V_{0.9}$) to the autonomous
oscillations that emerge at $V_{Hopf}$.

As shown in Fig.~\ref{fig:dynamic}d, the dynamic range is a rather
robust feature of our excitable circuit: it changes little as the
noise intensity is varied, regardless of the time scale at which it
operates. In both cases, $\Delta \simeq 6$~dB, which is closer to the
values obtained experimentally ($\Delta \simeq 10$~dB for olfactory
sensory neurons~\cite{Rospars00}, $\Delta \simeq 14$~dB for retinal
ganglion cells~\cite{Deans02,Furtado06}) than results obtained
theoretically for discrete models of excitable elements ($\Delta
\simeq 14$~dB in~\cite{Furtado06} and $\Delta \simeq 19$~dB
in~\cite{Assis08}).

\section{\label{conclusions}Concluding remarks}

In summary, we have presented an excitable electronic circuit whose
simplicity allows for scalability and accurate mathematical
modeling. Its dynamical equations lead to time series which
quantitatively reproduce experimental results without fitting
parameters. 

In addition, we have shown that the introduction of noise from a
simple analog noise generator at the input of the circuit produces
variable spike trains. The statistics of the interspike intervals is
shown to exhibit coherence resonance. Furthermore, by analyzing long
time series under low noise intensity, the spike trains were shown to
behave as a Poisson process, like some biological neurons. 

In the excitable regime, with fixed noise amplitude, the firing rate
response of the system to a $V_{DC}$ input -- the stimulus -- was
shown to have a dynamic range of about 6~dB, which is also comparable
to some biological sensory neurons. Together with its scalability,
these properties render the system a potential building block for
artificial sensors based on collective properties of excitable media.

\section{Acknowledgments}
%\begin{acknowledgments}
BNSM, MC and JRRL acknowledge financial support from Brazilian
agencies CNPq, FACEPE, CAPES and special programs PRONEX, PRONEM and
INCEMAQ. GBM acknowledges support from NIH. It is a pleasure to thank
Hugo L. D. S. Cavalcante for enlightening discussions during the
preparation of this work, as well as Marcos Nascimento for technical
support.
%\end{acknowledgments}

\bibliography{electronic-circuit-paper,copelli}
\bibliographystyle{ws-ijbc}
\end{document}